 \documentclass[twocolumn,showpacs,preprintnumbers,amsmath,amssymb]{revtex4}

\usepackage{graphicx}
\usepackage{dcolumn}
\usepackage{bm}

\begin{document}

\title{Theory of stripe domains
in magnetic shape memory alloys}
\author{N. S. Kiselev$^{1,2}$}
\thanks
{Corresponding author 
 \\IFW Dresden,
 Postfach 270116, D--01171 Dresden, Germany.\\ 
 Tel.: +49-351-4659-542; Fax: +49-351-4659-537
}

\email{m.kyselov@ifw-dresden.de}
\author{I. E. Dragunov$^2$}
\author{A. T. Onisan$^1$}
\author{U. K. R\"o\ss ler$^1$}
\email{u.roessler@ifw-dresden.de}
\author{ A. N.\ Bogdanov$^1$}

\affiliation{$^1$ IFW Dresden, Postfach 270116, D-01171 Dresden, Germany}

\affiliation{$^2$ Donetsk Institute for Physics and Technology, 
 83114 Donetsk, Ukraine}

\begin{abstract}
The evolution of multivariant patterns 
in thin plates of magnetic shape memory materials
with an applied magnetic field 
was studied theoretically.
A geometrical domain-model is considered 
composed of straight stripe-like martensite variants
with constant internal magnetization (high anisotropy limit) 
and magnetic domain wall orientation fixed by the 
twin boundaries.
Through integral transforms of the demagnetization energy, 
the micromagnetic energy is
cast into a form convenient 
for direct numerical evaluation 
and analytical calculations.
The equilibrium geometrical parameters
of multivariant patterns with straight and 
oblique twin boundaries
have been derived as functions of the applied field
and the material parameters of a plate.
It is shown that the oblique multivariant states exist only
in plates with thicknesses $L$ larger than a certain critical
value $L_0$. 
In samples with $ L < L_0$ a magnetic-field-driven
transformation occurs directly between single variant states.
\end{abstract}

\maketitle
\section{Introduction}
\label{intro}

In ferromagnetic shape memory alloys 
like Ni-Mn-Ga Heusler alloys
the high magnetocrystalline
anisotropy fixes the magnetization 
along the easy magnetization axis of
martensite variants \cite{Pan00}.
%

%
This ``one-to-one correspondence''
between magnetic domains and the martensite
variants allows to rearrange the crystallographic 
variants by applying a magnetic field
\cite{Pan00,Heczko01}. 
Hence, magnetic energy determines 
the evolution of the multivariant states 
in magnetic fields. 
The martensitic microstructure 
can be described by adapting 
the phenomenological theory of 
magnetic domains \cite{JMMM03,Hubert98}.
A detailed understanding of 
this coupling between microstructure
and magnetic domain structure can 
be achieved for suitably simple 
geometrical systems.

Here, we develop a theory 
of multivariant stripes in thin plates.
For these stripe structure we derive
analytical expressions for demagnetization energies 
and complete phase diagrams for equilibrium structures.

\section{Micromagnetic energy and equations}
\label{sec:1}
As a model we consider multivariant
states in a layer of thickness $L$
with surface normal along $z$
and infinite extension 
in $x$ and $y$ direction.
Commonly observed patterns include
lamellar microstructures built 
from two coexisting single-variant states 
with oblique or straight interfaces,
i.e. the angle between easy-magnetization
axis and $z$ is equal to $0$ or $\pi/4$, respectively  
(Fig. 1 a, b). The orientation of these interfaces
with 90-degree magnetic walls are fixed by the crystallography
of the twin-boundaries between the martensite variants.
Within individual variants stripe domains with 180-degree
magnetic domain walls can occur
(Fig. 1, c) 
\cite{Murakami03,Heczko04,Chernenko06,Lai07}.

The patterns in Fig. 1 a, b 
on the one hand satisfy the elastic 
compatibility between the tetragonal variants and, 
on the other hand, comply with a common property of 
magnetic domains by avoiding uncompensated magnetostatic 
charges on the domain boundaries.
Here, we assume that 
internal magnetic charges, that may arise 
at the twin-boundary \cite{Lai07}, are absent or 
can be neglected as the period of 180-degree
domain structures within variants remain small.
Further, we assume that the uniaxial anisotropy 
is much stronger than the applied fields.
Thus, deviations of the magnetization from the easy axis
within the variants can be neglected.
As sketched in Fig. 1 a,  
the magnetization in the  microsctructure
with oblique twin interfaces
can be reduced to a pattern with alternating 
magnetization $ \mathbf{m} = \pm(M/2)\mathbf{z}$ 
perpendicular to the plate surface 
and an effective bias field $ H = H_z + 2 \pi M$.
For the pattern with straight interfaces in Fig. 1 b, 
the alternating magnetization 
$\mathbf{m} = \pm (M/\sqrt{2})\mathbf{z}$, 
and the bias field is $H = H_z$.
Thus, the patterns in Fig. 1
are reduced to models of stripe domains
with alternating magnetization $\mathbf{m}$
perpendicular to the surface 
and \textit{straight} and \textit{oblique} 
domain walls in a bias field $H$.
The model with straight domain walls 
has been investigated in details for 
magnetic films (see e. g. \cite{FTT80,Hubert98})
and the solutions for the micromagnetic problems
can be applied directly for magnetic shape memory films.
The model with oblique interfaces is the main subject 
of our investigations in this paper.

\begin{figure}[tbh]
\includegraphics[width=6.5cm] {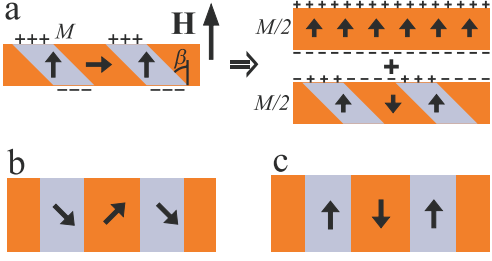}
\caption{Multivariant patterns
consisting of two different
tetragonal martensites, 
(a) $\beta = \pi/4$,
(b) $\beta = 0$, and stripe domains
within one variant (c).}
\label{Fig1}
\end{figure}
The reduced energy density $w = W/(2 \pi m^2$)
for the oblique stripes 
can be written
\begin{eqnarray}
 w = \frac{2\sqrt{2}p}{\pi}\frac{L_c}{L}-\frac{H}{2\pi m} q
+ \underbrace{q^2 +\frac{4}{\pi^2 p}\Xi(p,q,\pi/4)}_{w_d},
\label{energy1}
\end{eqnarray}
where $p = 2\pi L/D$, $ q = (d_{+}-d_{-})/D$, and
$D = d_{+}+d_{-}$ is the period of the domain structure,
$d_{\pm}$ are the widths of domains polarized in
the directions parallel (+) and antiparallel (-)
to the field.
The domain wall energy (the first term in Eq. (\ref{energy1})
includes the characteristic length
$L_c =  \sigma/(4\pi m^2)$ where $\sigma$ 
is the domain wall energy density
which describes the balance between the 
domain wall and stray field energies.
The stray field energy $w_d$ 
is given by a function $\Xi(p,q, \pi/4)$
that includes an infinite sum,
\begin{eqnarray}
\Xi(p,q, \beta) = \sum_{n=1}^{\infty}\frac{\left(
1-\left(-1\right)^n\cos\left(\pi n q\right)\right)}{n^3}\times \\
\nonumber
\left[1-\cos\left(n p \tan \beta \right)\exp\left(-np\right)\right] .
\label{stray2}
\end{eqnarray}

With the help of the integral transformation
introduced in \cite{FTT80} 
the infinite sum in Eq.~(\ref{stray2})
is transformed into integrals on the interval [0,1].
Then, the energy (\ref{energy1}) can be written
\begin{equation}
w =1-
\frac{4p}{\pi^2}\int\limits_0^1{( 1-t)\arctan \left[f(\xi,q)\right]dt}+
\frac{p}{\pi}\frac{L_0}{L}-2hq\,,
	\label{energy2}
\end{equation}
where $h = H/(4 \pi m)$, $L_0= 2\sqrt2 L_c$, $\xi = pt$,

\(
f(\xi,q)=2 \cos^2(\pi q/2)\sinh \xi \sin \xi \times 
\)

\(
[g(\xi,q)-2 \cos^2(\pi q/2)(\cosh \xi \cos \xi +\cos \pi q)]^{-1} \,\, \)and

\(g(\xi,q)= (\cosh \xi \cos \xi + \cos \pi q)^2 +\sinh^2 \xi \sin^2 \xi\).

For straight stripes ( $\beta=0$ )
the integral transformation of sum (\ref{stray2})
the following expression gives
the system  energy  \cite{FTT80}
\begin{eqnarray}
\label{energyST}
w=\!1\!-\!\frac{2 p}{\pi^2}\!\!\int\limits_0^1\!(1\!-\!t)\ln\!\! 
\left[1\!+\! \frac{ \cos^2 \left( \pi q/2\right) }
{ \sinh^2 \left(p t/2\right)}\!\right]\!\!dt\!+\!
\frac{p}{\pi}\frac{L_c}{L}\!-\!2hq\,.
\end{eqnarray}
Minimization of Eqs.~(\ref{energy2}) or 
(\ref{energyST}) with respect to
the internal parameters $p$, $q$
gives the equilibrium values for
the domain sizes ($d_{+}$, $d_{-}$).

\section{Results}

\subsection{Transition into the single variant states}

%
The difference $\Delta w = w- w_0$
between the system energy $w$ and that
of the homogeneously magnetized layer,
$\,\,w_0 = 1 - H/(4 \pi m)$, can be written
in the following form

\begin{eqnarray}
\nonumber
\Delta w =  -\frac{8 p}{\pi^2} 
\cos^2 \left(\frac{\pi q}{2}\right)
\int\limits_0^{1}\frac{(1-t)}{g(\xi,q)}\times \;\;\;\;\;\;\;\;  \\
\left[ \frac{\xi G(\xi,q)}{\cosh \xi -\cos \xi} -
\sinh \xi \sin \xi 
\frac{ \pi (1- q)}{\tan[\pi(1-q)/2]} \right]dt\,,
\label{deltaw}
\end{eqnarray}

where
$G (\xi,q) = (\sinh \xi - \sin \xi)( \cos \pi q + \cosh \xi \cos \xi)-
\sinh \xi \sin \xi (\sinh \xi + \sin \xi)$.

\begin{figure}
		\includegraphics [width=6.0cm] {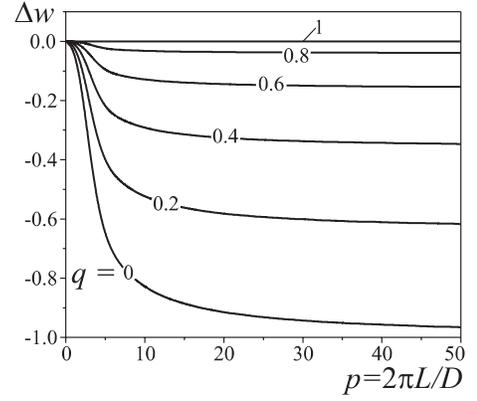}
%
\caption{ Energy difference $\Delta w (p)$ 
between stripe and single variant 
equilibrium states as function of the reduced
inverse stripe period $p$ and for different values 
of the reduced difference $q$ between widths of 
up and down domains.
The equilibrium energy of the multidomain 
states is lower than that of the homogeneous
state.
}
\label{energyprofiles}
\end{figure}

\begin{figure}
		\includegraphics[width=7.5cm] {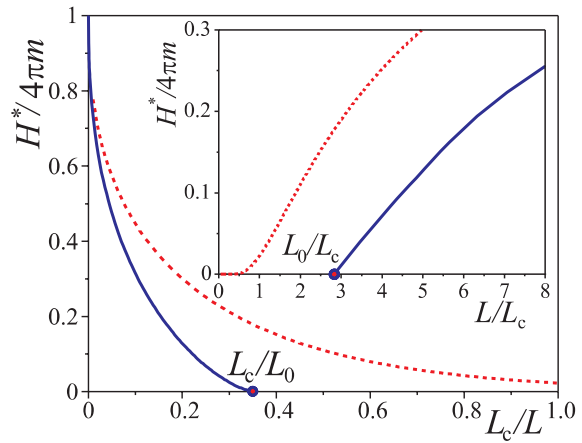}
\caption{Phase diagram in terms of the inverse
reduced plate thickness $L_c/L$ 
and the reduced magnetic field $H/(4 \pi m)$ 
perpendicular to the plate 
for oblique (solid) and straight (dashed) stripes.
The oblique domains do not exist below
the critical thickness $L_0 = 2\sqrt{2}L_c$.
Inset shows the functions $H^* (L/L_c)$
close to the critical thickness.
}
\label{trans}
\end{figure}

The energy difference $\Delta w (p,q)$,
Eq.~(\ref{deltaw}),
is \textit{negative} for all  $ p > 0$
and $0 < q <1$,  and reaches zero only for $p=0$, $q =1$
(Fig. \ref{energyprofiles}).
Similar relations are also true for straight stripes \cite{FTT80}.
For both types of systems,
stripes with finite widths
have 
\textit{always} lower energy than 
a single variant state.
The stripes transform into a single
variant state continuously by
unlimited growth of the period $D$.
%
%
Near the transition into the single variant 
state $ p \rightarrow 0$, and, thus,
the function $f(\xi)$ in Eq.~(\ref{energy2})
can be expanded with respect to
the small parameter $\xi$, 
and the energy (\ref{energy2})
can be evaluated in terms of elementary
functions
\begin{equation}
w =1- \cos{ \left( \frac{\pi q}{2}\right)}
\left[ \frac{2}{\pi u}\frac{L_0}{L}-\dfrac{2}{\pi^2}\Omega(u)\right]
-2hq\,,
\label{energy3}
\end{equation}
where 

\(
\Omega (u)= \dfrac u2 \ln{\left(1+\dfrac{4}{u^4}\right)}
+\ln{\left[\dfrac{(1+(u+1)^2)}{(1+(u-1)^2)}\right]} -\upsilon (u)
\),

$u =2\cos{(\pi q/2)}/p$
, and $\upsilon (u) =2(1+u^{-1})\arctan (1+u^{-1})-
2(1-u^{-1})\arctan(1-u^{-1})$.
Minimizing $w$ from Eq.~(\ref{energy3}) 
with respect to $u$ and $q$ 
yields equations for equilibrium values of $p$, $q$ and $u$
\begin{eqnarray}
\nonumber
p(u)=\frac{2}{u}\sqrt{1-\left(\frac{h}{h^{*}} \right)^2},\\
\nonumber
\quad
q(u)=\frac{2}{\pi}\arcsin\left(\frac{h}{h^{*}} \right)
\label{sol},\\
\quad
2 \pi L_0 \Upsilon (u) = L,
\end{eqnarray}
where 
\begin{equation}
  \Upsilon (u) = \left[u^2\ln\left(1+\frac{4}{u^4}\right)+
    4\arctan{\left( \frac{u^2}{2}\right)} \right]^{-1},
\label{u}
\end{equation}
\begin{eqnarray}
\nonumber
h^* (u) = \frac{1}{2 \pi} \left[2 \arctan{\left(\frac{2 u}{u^2-2}\right)}-\right. \\
\left. u \ln\left(1+\frac{4}{u^4}\right)-
\ln \left(\frac{1+(u+1)^2}{1+(u-1)^2}\right)\right].
\label{hcrit}
\end{eqnarray}
Eqs.~(\ref{sol}) express 
the equilibrium values of $p$ and $q$
in parametrized form 
as functions of the applied field  $h$ and the plate
thickness $L$.
According to Eqs.~(\ref{sol}) the parameter $h^*$
from Eq.~(\ref{hcrit}) defines the transition field 
into the saturated state ($p=0$, $q =1$).
Note that for $h=h^*$ the parameter $u$ transforms
into  $u(h^*) =  d_{-}/L$.
This means that the transtion into the
saturated state takes place by 
an infinite expansion of the domains 
with the magnetization parallel to the field, 
($ d_{+} \rightarrow \infty$), while domains with
the antiparallel magnetization ($d_{-}$) 
keep a finite size. 
The transition line $h^*$  plotted as 
function of $L_c/L$ in Fig.~\ref{trans}
separates the regions of multivariant
and single variant states.
The equation $h^* =0$ defines the
\textit{critical thickness}
$L_0 = 2\sqrt{2}L_c$.
For $L < L_0$ the oblique domains do
not exist, while straight domains
(theoretically) exist for any thickness
\cite{FTT80,Hubert98} (Fig.~\ref{trans}). 

\subsection{Domain evolution in the magnetic field. 
Magnetization curves}
\begin{figure}[tbh]
\includegraphics[width=7.0cm] {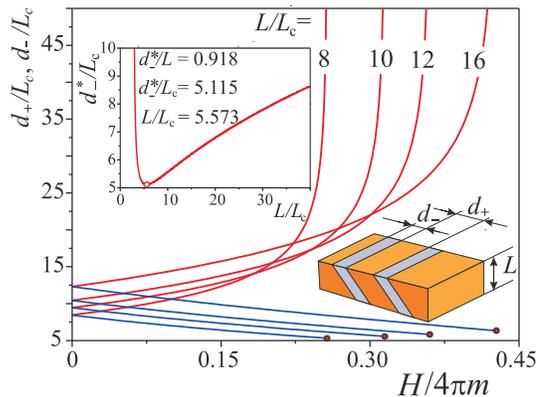}
%
\caption{
The equilibrium stripe domain sizes
$d_{+}/L_c$ (upper set of curves) $d_{-}/L_c$ 
(lower set of curves) as  functions of the applied 
magnetic field for different
layer thicknesses.
Inset: dependence of the size
for the minority domain 
on layer thickness at the transition. 
}
\label{solutions}
\end{figure}

Typical solutions for domain sizes ($d_{+}, d_{-}$) 
are presented in Fig.~\ref{solutions}.
At low fields, the domain sizes change only slowly.
Exponential growth of the size $d_{+}$ for 
the domains with magnetization in field direction
sets in close to the saturation field.
The minority domain size $d_{-}$ gradually 
decreases in the increasing magnetic field 
but remains finite at the transition field.
\begin{figure}[tbh]
\includegraphics[width=8.5 cm] {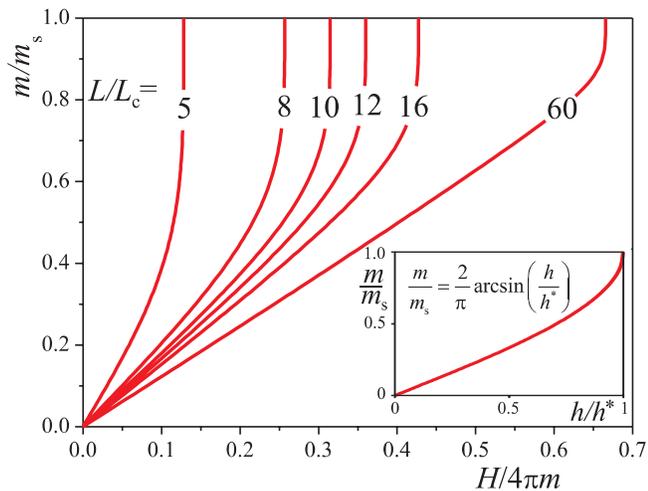}
\caption{ Magnetization curves for different
values of the reduced thickness $L/L_c$.
Inset shows the ``universal'' character
of the magnetization curves for large
domains.
}
\label{curve}
\end{figure}
%
The equilibrium magnetization curves
in Fig.~\ref{curve} 
display different 
behavior of the magnetization reversal 
in ``thin'' ($L \geq L_c$) and ``thick''
($L \gg L_c$) plates.
The former are characterized by
low values of the saturation
field $h^* \ll 1$ and steep 
magnetization curves.
In accordance with the phase theory
the latter have linear magnetization
curves $m/m_s = h$ in a broad range
of the applied field. 

\begin{figure}[tbh]
		\includegraphics[width=8.5 cm] {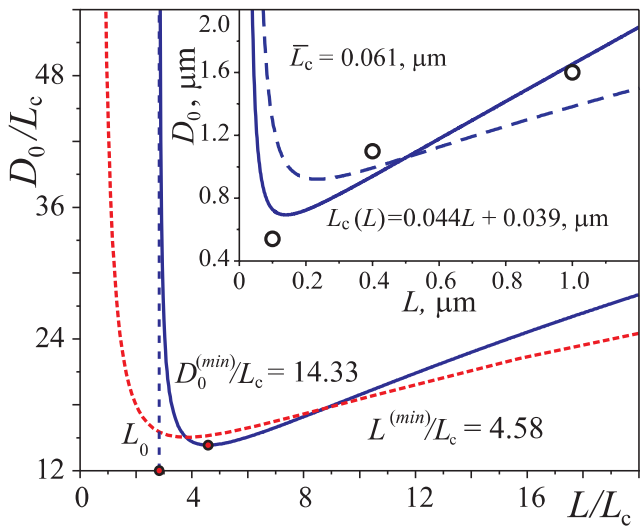}
%
\caption{
The equilibrium period
$D_{0}/L_c$ as a function of the plate 
thickness $L/L_c$ at zero field 
(solid line) in comparison with 
a correponding line (dashed) 
for straight stripes.
Inset shows $D(L)$ functions
for a thin layer of Ni$_{51.4}$Mn$_{28.3}$Ga$-{20.3}$
investigated in Ref. \cite{Chernenko06} 
corresponding to the straight stripe structure Fig.~\ref{Fig1}, b.
The dashed curve is plotted for the
averaged characteristic length
$\bar{L}_c$, the solid line incorporates
effects connected with
the indicated variation of $L_c$ with $L$.
}
\label{solutions0}
\end{figure}

%
In various aspects,
we find that the solutions 
for the oblique stripes have 
a \textit{universal} character 
that may be useful for analysis of experimental data.
The solutions for $q$ in Eq.~(\ref{sol}) 
give a universal magnetization curve for systems 
with large domains $m/m_s = (2/\pi) \arcsin(h/h^*)$
(Inset in Fig.~\ref{curve}).
The solutions
for the equilibrium values of $p$
%
\(
(8/\pi)\int_0^1 t 
\arctan{\left[ \sin pt/\sinh pt\right]}dt=L_0/L
\)
at zero field ( $q=0$) 
%
yield a specific dependence of zero-field
period $D_0$ on the layer thickness
(Fig.~\ref{solutions0}). 

The equation for $p$ corresponding
to the minimum value of $D_0$,
\(
\arctan\left[ \sin p/\sinh p\right]=
\int_0^1 t \arctan{\left[ \sin pt/\sinh pt\right]}dt
\),
%
includes no material parameters and has the solution
$p_{min}=2.01$. 
This yields 
$L^{(min)}/L_{c} = 4.58$, $D_0^{(min)}/L_{c}  =14.33$ and the 
ratio $D_0^{(min)}/ L^{(min)}= 3.13$.
Corresponding values for straight domains 
are $\tilde{p}_{min} = 1.595$,
$\tilde{L}^{(min)}/L_c = 3.8178$, 
$\tilde{D}_0^{(min)}/L_c = 14.5739$, 
$\tilde{D}_0^{(min)}/ \tilde{L}^{(min)}= 3.8176$
\cite{condmat06}.
The minority domain size 
, $d_{-}^*/L_c (L/L_c)$, 
at the transition field
has a similar minimum (Inset in Fig.~\ref{solutions}).
The equation for $ x = \min(d_{-}^*/L)$ 
%
$
4\arctan\left(x^2/2\right)-x^2\ln\left(1+4/x^4\right)=0
$
%
has the solution  $x=0.918$, and yields
the following values of the parameters
in this point:
$L/L_{c}=5.573$,
$d_{-}^*/L_{c}=5.115$.

The point $D_0^{(min)}$ marks the transition region
between a functional dependence $ D_0 \propto sqrt{L}$
characteristic for thick plates ($ D_0 \ll L$)
(\textit{Kittel} law) to a exponential growth of 
the period with a decreasing plate thickness,
$D_0 \propto \exp(-\pi L_c/L)$\cite{FTT80}
in thin layers ($ L \ge L_c$). 
This transitional region has been reached in thin
polycrystalline films of Ni$_{51.4}$Mn$_{28.3}$Ga$-{20.3}$
with easy-axes dominantly at 45-degree from the layer normal
direction \cite{Chernenko06}.
Inset in  Fig.~\ref{solutions0} shows 
the calculated functions $D_0 (L)$
with material parameter $L_c$ adjusted 
to the experimental data of Ref.~\cite{Chernenko06}.

\section{Conclusions}
The present theory enables 
rigorous calculations of equilibrium 
structures in thin ferromagnetic martensite 
plates with two variants. 
It is clear that real magnetization processes
in magnetic shape-memory processes will generally 
be hysteretic. However, the equilibrium states
could be reached, e.g., free-standing films or
thin free single-crystal platelets with very
mobile twin-boundaries.
The analytical expressions for 
the energy of a stripe system 
Eqs.~(\ref{energy1})--({energyST}) 
are applicable also for non-equilibrium states.
We find that oblique stripe structures
are possible only above a certain critical 
thickness, while straight stripe structures 
have been known to exist for arbitrary 
dimensions of layers with
perpendicular magnetic anisotropy.
The theoretical method can 
be extended to treat twinned martensite 
plates with other angle between easy 
magnetization and layer normal direction.

\section*{Acknowledgements}
The authors are grateful to J. McCord,
Y. W. Lai, V. Neu, and R. Sch\"afer 
for helpful discussions.
N.S.K., A.N.B.  \ thank H.\ Eschrig for
hospitality at IFW Dresden. 
We gratefully acknowledge support by 
Deutsche Forschungsgmeinschaft (SPP 1239 project A8).
%

%
%

\end{document}